# POMPALANMIŞ SU TABANLI ENERJİ DEPOLAMA ÜNİTESİ KULLANIMINA VE DİNAMİK FİYATLANDIRMAYA DAYANAN BİR KAZANÇ OPTİMİZASYONU YAKLAŞIMI

Akın TAŞCIKARAOĞLU[1*], Ozan ERDİNÇ[2]

[1]Muğla Sıtkı Koçman Üniversitesi, Mühendislik Fakültesi, Elektrik-Elektronik Mühendisliği Bölümü, 48000 Kötekli yerleşkesi, Muğla, e-posta : akintascikaraoglu@mu.edu.tr,
ORCID No : http://orcid.org/0000-0001-8696-6516

[2]Yıldız Teknik Üniversitesi, Elektrik Elektronik Fakültesi, Elektrik Mühendisliği Bölümü C Blok, Davutpaşa Mah., Davutpaşa Caddesi 34220 Esenler- İstanbul, e-posta : oerdinc@yildiz.edu.tr
ORCID No : http://orcid.org/0000-0002-6854-1853

**Pompalanmış Su Tabanlı Enerji Depolama Ünitesi Kullanımına ve Dinamik Fiyatlandırmaya Dayanan Bir Kazanç Optimizasyonu Yaklaşımı**

**Öz**

*Bu çalışmada, değişken ve süreksiz enerji üretimine sahip olan rüzgar çiftliklerinden, gün öncesi ve dengeleme piyasalarının mevcut olduğu bir güç sisteminde en yüksek ekonomik kazancı elde edebilmek amacıyla bir optimizasyon problemi önerilmektedir. Pompalanmış su tabanlı bir enerji depolama ünitesinin rüzgar çiftliği ile birlikte kullanılmasına dayanan bu yöntem, enerji piyasalarındaki fiyat değişimden faydalanarak güç santralinden elde edilen kazancı arttırmaktadır ve aynı zamanda santralin bağlı olduğu güç sistemindeki tepe yük talebinin bir kısmını karşılayarak sisteme destek olmaktadır. Önerilen yaklaşımın etkinliğini inceleyebilmek amacıyla gerçek rüzgar ve fiyat verilerinden faydalanılarak detaylı benzetim çalışmaları gerçekleştirilmiştir ve sonuçlar depolama ünitesinin olmadığı ve önerilen fiyat bazlı enerji yönetimi yaklaşımının kullanılmadığı çeşitli durumlar için elde edilen sonuçlar ile karşılaştırılmıştır. Sonuç olarak, pompalanmış su tabanlı enerji depolama ünitelerinin yüksek güçlerde etkin bir şekilde kullanılabilecek bir enerji depolama sistemi olduğu ve bu sistemin ekonomik açıdan en verimli şekilde kullanılabilmesinde önerilen optimizasyon probleminin oldukça başarılı olduğu gösterilmiştir.*

**Anahtar Kelimeler :** *Pompa depolamalı hidroelektrik santral, rüzgar türbini, elektrik enerji depolama sistemleri, elektrik piyasaları.*

**A Profit Optimization Approach Based on the Use of Pumped-Hydro Energy Storage Unit and Dynamic Pricing**

**Abstract**

*In this study, an optimization problem is proposed in order to obtain the maximum economic benefit from wind farms with variable and intermittent energy generation in the day ahead and balancing electricity markets. This method, which is based on the use of pumped-hydro energy storage unit and wind farm together, increases the profit from the power plant by taking advantage of the price changes in the markets and at the same time supports the power system by supplying a portion of the peak load demand in the system to which the plant is connected. With the objective of examining the effectiveness of the proposed method, detailed simulation studies are carried out by making use of actual wind and price data, and the results are compared to those obtained for the various cases in which the storage unit is not available and/or the proposed price-based energy management method is not applied. As a consequence, it is demonstrated that the pumped-hydro energy storage units are the storage systems capable of being used effectively for high-power levels and that the proposed optimization problem is quite successful in the cost-effective implementation of these systems.*

**Keywords :** Pumped-storage hydroelectric plant, wind turbine, electric energy storage systems, electricity markets.

---

* Sorumlu Yazar; 0 252 211 5765

# 1. Adlar Dizini

## 1.1. Kümeler:

| | |
|---|---|
| $s$ | Senaryo |
| $t$ | Zaman [saat] |

## 1.2. Parametreler:

| | |
|---|---|
| $N$ | Yeterince büyük pozitif sabit |
| $P_t^{planlanan,göp}$ | Gün öncesi piyasasında $t$ anında şebekeye verilmesi planlanan güç [MW] |
| $P_{s,t}^{rüzgar}$ | Rüzgar çiftliğinin $t$ anında ürettiği güç [MW] |
| $Q^{maks,ps}$ | Türbin/pompa modunda maksimum su akışı [Hm³/h] |
| $V^{ALT,ps,başl}$ | Alt rezervuarın başlangıçtaki su miktarı [Hm³] |
| $V^{ALT,ps,maks}$ | Alt rezervuarın maksimum su miktarı [Hm³] |
| $V^{ALT,ps,min}$ | Alt rezervuarın minimum su miktarı [Hm³] |
| $v_t^{buhar}$ | Buharlaşma nedeniyle rezervuar su miktarında meydana gelen azalma [Hm³] |
| $V^{ÜST,ps,başl}$ | Üst rezervuarın başlangıçtaki su miktarı [Hm³] |
| $V^{ÜST,ps,maks}$ | Üst rezervuarın maksimum su miktarı [Hm³] |
| $V^{ÜST,ps,min}$ | Üst rezervuarın minimum su miktarı [Hm³] |
| $v_t^{yağmur}$ | Yağmur nedeniyle rezervuar su miktarında meydana gelen artış [Hm³] |
| $\lambda_{s,t}^{alınan,dp}$ | Dengeleme piyasasında $t$ anında satın alınan enerjinin fiyatı [TL/MWh] |
| $\lambda_t^{göp}$ | Gün öncesi piyasasında $t$ anında satılan enerjinin fiyatı [TL/MWh] |
| $\lambda_{s,t}^{satılan,dp}$ | Dengeleme piyasasında $t$ anında satılan enerjinin fiyatı [TL/MWh] |
| $\sigma^{salınan,ps}$ | Su akışından serbest bırakılan güce çevrim katsayısı [MW/( Hm³/h)] |
| $\sigma^{pompalanan,ps}$ | Su akışından pompalanan güce çevrim katsayısı [MW/( Hm³/h)] |
| $\Delta T_{dp}$ | Dengeleme piyasası için zaman çözünürlüğü [h] |
| $\Delta T_{göp}$ | Gün öncesi piyasası için zaman çözünürlüğü [h] |

## 1.3. Değişkenler:

| | |
|---|---|
| $P_{s,t}^{alınan,dp}$ | Dengeleme piyasasından $s$ senaryosunda $t$ anında alınan güç [MW] |
| $P_{s,t}^{pompalanan,ps}$ | Üst rezervuara $s$ senaryosunda $t$ anında pompalanan su miktarı için pompalanmış su tabanlı enerji depolama ünitesine gönderilen güç [MW] |
| $P_{s,t}^{salınan,ps}$ | Alt rezervuara $s$ senaryosunda $t$ anında serbest bırakılan su miktarı için pompalanmış su tabanlı enerji depolama ünitesinden elde edilen güç [MW] |
| $P_{s,t}^{satılan,dp}$ | Dengeleme piyasasına $s$ senaryosunda $t$ anında satılan güç [MW] |
| $q_{s,t}^{pompalanan,ps}$ | Alt rezervuardan $s$ senaryosunda $t$ anında pompalanan suyun akış hızı [Hm³/h] |
| $q_{s,t}^{salınan,ps}$ | Üst rezervuardan $s$ senaryosunda $t$ anında serbest bırakılan suyun akış hızı [Hm³/h] |
| $u_{s,t}^{dp1}$ | Dengeleme piyasasında güç değişimi için ikili değişken |
| $u_{s,t}^{dp2}$ | Dengeleme piyasasında arbitraj için ikili değişken |
| $u_{s,t}^{ps}$ | Pompa depolamalı santralin serbest bırakılan ve pompalanan güç değişimi için ikili değişken |
| $v_{s,t}^{ALT,ps}$ | Alt rezervuarın $s$ senaryosunda $t$ anındaki su miktarı [Hm³] |
| $v_{s,t}^{ÜST,ps}$ | Üst rezervuarın $s$ senaryosunda $t$ anındaki su miktarı [Hm³] |

# 2. Giriş

Artan elektrik enerjisi talepleri ve bu taleplerin konvansiyonel enerji kaynakları ile karşılanması sonucunda ortaya çıkan çevresel kaygılarla birlikte yenilenebilir enerji kaynakları dünya genelinde artan bir şekilde yaygınlaşmaya devam etmektedir. Bu kaynaklar içerisinde rüzgar türbinleri ve rüzgar çiftlikleri, diğer kaynaklara oranla daha yüksek güç üretimi kapasitelerine ve daha düşük işletme maliyetlerine sahip olduklarından günümüzde en çok kullanılan yenilebilir enerji kaynağı konumundadırlar. 2016 yılı sonu raporlarına göre rüzgar enerjisi son on yılda ortalama %21'lik bir artış ile tüm dünyada yaklaşık 500 GW'a ulaşmıştır (Global Wind Energy Council, 2016).

Kullanılmaya başladığı ilk zamanlarda sınırlı kurulu güç kapasitesine sahip olduğundan şebeke üzerinde olumsuz bir etkisi gözlemlenmeyen rüzgar enerjisinin şebeke içerisindeki oranının artmasıyla birlikte çeşitli işletimsel sorunlar da ortaya çıkmıştır. Bu sorunların en önemli nedeni, rüzgar türbinlerinden elde edilen enerjinin süreksiz ve değişken yapıda olmasıdır. Rüzgar türbinlerinde üretilen güç miktarının çok büyük oranda rüzgar hızına bağlı olması, konvansiyonel santrallerden elde edilen enerjinin kontrolünde kullanılan yöntemlerin rüzgar enerjisi için uygulanmasını oldukça zorlaştırmaktadır.

Rüzgar enerjisinin belirtilen nedenlerden dolayı kontrol edilememesi, dağıtım ve hatta iletim hatlarındaki üretim ve tüketim dengesinin bozulmasına sebep olabilmektedir. Enerji arz ve talebindeki bu uyumsuzluklar gerilim ve frekans değerlerinde sapmalara yol açmaktadır. Şebekelerdeki bu olumsuz etkileri en aza indirebilmek amacıyla literatürde ve gerçek uygulamalarda kullanılan yöntemler üç farklı başlık altında toplanabilir: (i) Gelişmiş tahmin yöntemleri kullanarak rüzgar enerjisindeki belirsizlikleri azaltmak, (ii) talep cevabı stratejilerini kullanarak yük tarafında yapılacak değişiklikler ile üretim-tüketim arasındaki dengeyi korumaya çalışmak ve (iii) rüzgar türbinlerini bir enerji depolama sistemi ile birlikte kullanmak.

Rüzgar türbinlerinin neden olabileceği olası sorunları azaltarak türbinlerden elde edilebilecek faydaları en yüksek seviyeye çıkarmayı hedefleyen ve yukarıda bahsedilen yöntemlerden ilki olan rüzgar hızını ve gücünü tahmin etmeye çalışmak, özellikle birkaç güne kadar olan kısa süreler için oldukça olumlu sonuçlar vermektedir (Tascikaraoglu, 2018b). Ancak bu yöntemler yapıları gereği her zaman bir belirsizlik içerdiklerinden sadece tahminlere güvenmek çoğu zaman yeterli olmamaktadır. Verilen ikinci yöntem ise akıllı şebeke teknolojilerinin yaygınlaşması ile birlikte özellikle son yıllarda geniş bir uygulama alanı bulmuştur. Temel olarak son kullanıcılara ait yüklerin, enerji üretiminin daha fazla ve/veya tüketiminin daha az olduğu zamanlara ötelenmesine dayanan bu yaklaşım, daha verimli bir işletim elde edebilmek amacıyla çoğu zaman rüzgar tahminleri ve yük tahminleri ile bir arada kullanılmaktadır (Paterakis vd., 2016; Elma vd., 2017). Bahsedilen üçüncü yöntem, rüzgar enerjisinin beklenen değerinden daha fazla veya daha az olması durumundaki olası problemlerin azaltılmasında ve belirli bir bölgedeki tepe enerji talebinin karşılanmasında ilk iki yönteme oranla daha etkilidir. Ancak buna rağmen bu yöntemin pratikteki ve literatürdeki uygulamaları ilk iki yönteme kıyasla çok daha sınırlıdır. Bunun en önemli nedeni, bir rüzgar çiftliğinin ürettiği MWh'ler mertebesindeki enerjinin konvansiyonel sistemlerle depolamasının ekonomik ve teknik olarak oldukça zor olmasıdır. Örneğin, en yaygın olarak kullanılan elektrik enerjisi depolama sistemi olan bataryalar, bu mertebelerdeki güçler için oldukça yüksek ilk yatırım, işletme ve bakım maliyetlerine sahip olmaktadır (Tascikaraoglu, 2018a). Bu amaçla, yenilenebilir enerjinin depolanması alanındaki çalışmalar, daha yüksek güç ve enerji kapasitelerine sahip olan pompalanmış su ve sıkıştırılmış hava tabanlı enerji depolama üniteleri üzerinde yoğunlaşmıştır.

Belirtilen enerji depolama teknolojilerinden pompalanmış su tabanlı enerji depolama üniteleri, doğal veya yapay bir su kaynağındaki suyun daha yüksek bir kotta bulunan su kaynağına aktarılması sayesinde elektrik enerjisinin potansiyel enerjiye dönüştürülmesi ve gerektiğinde bu potansiyel enerjinin tekrar elektrik enerjisine dönüştürülmesi prensibine dayanmaktadır. Literatürde farklı depolama sistemlerini bir arada ele alan çalışmalarda pompalanmış su tabanlı ünitelerin, özellikle sistemde yüksek seviyelerde rüzgar gücü bulunması durumunda, sıkıştırmalı hava tabanlı depolama ünitelerine ve diğer depolama sistemlerine oranla enerji maliyetini daha yüksek oranda azaltabildikleri belirtilmektedir (de Boer vd., 2014).

Literatürde çok sayıda çalışmada pompalanmış su tabanlı enerji depolama ünitelerinin çeşitli kullanım alanları ve potansiyel faydaları farklı açılardan incelenmiştir. Ma vd. (2014) şebekeden bağımsız bir ada için pompalanmış su tabanlı enerji depolama üniteleri ve bataryaları gerçek bir proje üzerinde maliyet ve çevrim ömrü gibi ekonomik açılardan karşılaştırmıştır ve pompalanmış su tabanlı ünitelerin daha verimli olduğu sonucuna varmıştır. Stoppato vd. (2014) ve Ma vd. (2015b) şebekeden bağımsız küçük bir bölge için pompalanmış su tabanlı ünitelerin fotovoltaik (PV) panellerle birlikte kullanılması durumunda elde edilebilecek ekonomik faydaları araştırmışlardır. Bu sistemlere bir rüzgar türbini eklenmesi durumunda ise maliyetlerin önemli derecede düştüğü (Ma vd., 2015a) ve (Petrakopoulou vd., 2016)'da gösterilmiştir. Enerji maliyetlerdeki bu değişimin en önemli nedeni, depolama sistemlerinin, PV panellere göre daha değişken bir güç üretimi karakteristiğine sahip olan rüzgar türbinleri ile birlikte kullanılmaya daha uygun olması olarak belirtilmiştir. Bu amaçla, şebekeden bağımsız adalar için pompa depolamalı hidroelektrik santrallerin bir rüzgar çiftliği ile birlikte kullanımının, toplam üretim içerisindeki yenilenebilir enerji oranını artırmak ve enerji maliyetini azaltmak açılarından sağladığı faydalar (Chen vd., 2016; Papaefthymiou ve Papathanassiou, 2014; Papaefthymiou vd., 2015; Pérez-Díaz ve Jiménez, 2016)'da incelenmiştir.

Yenilenebilir enerji kaynaklarından elde edilen enerjinin tüketimden fazla olması durumunda bu enerjinin depolanması ve sistemdeki enerji üretiminin talebi karşılamadığı zamanlarda depolanan enerjinin sisteme aktarılması amacıyla pompalanmış su tabanlı enerji depolama ünitelerinin şebekeden bağımsız sistemlerde kullanılması, yukarıda verilen örneklerden görülebileceği üzere özellikle enerji maliyetinin azaltılması açısından önemli faydalar sağlamaktadır. Ancak pompalanmış su tabanlı enerji depolama ünitelerinin şebekeden bağımsız sistemlerdeki faydaları genellikle ekonomik faydalarla sınırlı kalmaktadır. Bu santrallerden, ekonomik faydalarının yanı sıra birim yüklenme ve ekonomik dağıtım gibi çeşitli güç sistemi optimizasyonu işlemlerine yardımcı olmak amacıyla da faydalanabilmek için literatürde çok sayıda çalışma gerçekleştirilmiştir. Son yıllarda gerçekleştirilen çalışmalar arasında, Vieira vd. (2016) ve Ming vd. (2014) rüzgar gücünün belirsizliğini azaltarak birim yüklenme problemini optimize etmek amacıyla bir rüzgar çiftliği ile pompalanmış su tabanlı bir enerji depolama ünitesini ile birlikte kullanan bir yapı önermişlerdir. Olasılıksal birim yüklenme problemi için benzer pompalanmış su tabanlı ünite-rüzgar çiftliği hibrit sistemleri (Khodayar vd., 2013) ve (Bruninx vd., 2016)'da incelenmiştir ve aynı problem için sisteme talep cevabı yöntemlerinin de dahil edilmesi durumundaki sistemin toplam etkinliği (Kiran ve Kumari, 2016)'da araştırılmıştır. Bir rüzgar türbini ve pompalanmış su tabanlı ünitenin de yer aldığı bir hibrit sistemin optimal dağıtım probleminde kullanılması sonucunda yenilenebilir olmayan kaynaklardan çekilen gücün önemli oranda azaltıldığı (Kusakana, 2016)'da belirtilmiştir. Başka bir çalışmada, pompalanmış su tabanlı ünitelerin şebekeye frekans düşmeleri esnasında da önemli bir destek sağlayabileceği gösterilmiştir (Attya ve Hartkopf, 2015).

Yukarıda bahsedilen çalışmaların tamamında, rüzgar türbinlerinin ürettiği elektrik enerjisindeki belirsizliğinin azaltılmasında ve bu sayede güç sistemi işletimine destek sağlanmasında pompalanmış su tabanlı enerji depolama ünitelerinin etkin bir şekilde kullanılabileceği belirtilmektedir. Ancak bu çalışmalar güç sistemlerindeki değişken enerji fiyatlarını ve farklı enerji piyasalarını göz önüne almamaktadırlar. Foley vd. (2015) tarafından gerçekleştirilen ve pompalanmış su tabanlı ünitelerin rüzgar türbinleri ile birlikte kullanılmasını uzun bir dönem için inceleyen çalışmada da sonuç olarak verildiği gibi, pompalanmış su tabanlı ünitelerin enerji depolama ve enerjiyi şebekeye geri verme işlemlerinde farklı enerji fiyatları dikkate alınmadığı sürece, bu ünitelerin ilk yatırım maliyeti çoğunlukla elde edilen ekonomik kazancın toplamından daha fazla olmaktadır. Bu hususu dikkate alarak Sousa vd. (2014) pompalanmış su tabanlı üniteler ile rüzgar türbinlerinin bir arada kullanıldığı sistemlerde pompalanmış su tabanlı ünitelerin fiyat belirleyici (price maker) ve fiyat alıcı (price taker) olmaları durumlarını ayrı ayrı incelemişlerdir. Çalışma sonucunda şebeke içerisindeki en yüksek rüzgar oranının, bu iki fiyatlandırma durumu için tamamen farklı değerler aldığı belirtilmiştir. Benzer şekilde, pompalanmış su tabanlı ünite ve rüzgar türbini içeren bir enerji üretimi sistemi için enerji ve düzenleme piyasalarında kullanılmak üzere bir enerji fiyatlandırma stratejisi (Al-Swaiti vd., 2017) ve (Yıldız ve Şekkeli, 2016)'da önerilmiştir. Bir diğer çalışmada ise pompalanmış su tabanlı ünite-rüzgar türbini sistemi için en uygun fiyatlandırma stratejisi; gün öncesi piyasa, dengeleme piyasası ve fiziksel ikili anlaşmalar (physical bilateral contracts) dikkate alınarak belirlenmiştir (de la Nieta vd., 2016).

Önerilen çalışmada, öncelikle pompalanmış su tabanlı bir enerji depolama ünitesi ve bir rüzgar çiftliğinden oluşan bir sistem için (Foley vd., 2015)'de belirtildiği gibi şebeke alışverişlerinde farklı enerji fiyatlarının dikkate alınması gerektiği düşünülmüştür. Ayrıca, (Sousa vd., 2014)'te verilen sonuçlar göz önüne alınarak, şebekeye en yüksek rüzgar gücü girişini sağlamak amacıyla, önerilen sistemin enerji piyasasındaki fiyatlara göre şebekeye enerji sağladığı varsayılmıştır. Bu amaçla, (Al-Swaiti vd., 2017; Yıldız ve Şekkeli, 2016; de la Nieta vd., 2016)'da sunulan farklı enerji piyasalarının dikkate alınması durumundaki olumlu sonuçlar göz önüne alınarak, sistemin bağlı olduğu şebekedeki fiyatlandırmanın gün öncesi ve dengeleme piyasaları ile belirlendiği bir sistem oluşturulmuştur. Bahsedilen tüm bu çalışmalara ek olarak, oluşturulan enerji üretimi sistemi ve fiyatlandırma stratejileri için güç santrali sahibinin ekonomik kazancını en üst seviyeye çıkaracak ve aynı zamanda tepe yük taleplerinde şebekeye destek olacak bir optimizasyon problemi ortaya konmuştur. Sistemin etkinliği, gerçek rüzgar hızları ve enerji fiyatları dikkate alınarak test edilmiştir. Ayrıca, benzetim çalışmalarında farklı şartlar altında sonuçlar elde edebilmek amacıyla, rüzgar gücündeki ve enerji fiyatlarındaki belirsizlikler dikkate alınmıştır. Son olarak, önerilen stokastik optimizasyon problemi ile elde edilen sonuçlar, oluşturulan üç farklı duruma ait sonuçlar ile karşılaştırılmıştır.

Bu çalışmanın 3. bölümünde önerilen optimizasyon problemine ait matematiksel formülasyona yer verilmektedir. Önerilen yöntemin uygulanmasıyla elde edilen benzetim sonuçlarının farklı üç yönteme ait sonuçlarla karşılaştırılması 4. bölümde sunulmaktadır. Son bölümde ise çalışmaya ait temel bulgular ve sonuçlar verilmektedir.

## 3. Güç Santralinin Yapısı ve Optimizasyon Probleminin Formülasyonu

Çalışma kapsamında göz önüne alınan pompalanmış su tabanlı enerji depolama ünitesi ve rüzgar çiftliğinden oluşan güç santralinin yapısına ait açıklamalar ile önerilen optimizasyon problemine ait formülasyon aşağıda alt başlıklar halinde verilmektedir.

### 3.1. Önerilen Yapının Genel Şeması

Önerilen pompalanmış su tabanlı enerji depolama ünitesi ve rüzgar çiftliği tabanlı hibrit güç santralinin yapısı Şekil 1'de gösterilmektedir. İlgili santralin güç üretimi gün öncesi enerji piyasasında 24 saatlik zamana bağlı olarak her saatteki ilgili gün öncesi piyasa fiyatıyla birlikte planlanmaktadır. Dengeleme piyasasında ise 15 dakikalık periyotlarla ilgili santral sistem bileşenleri arasındaki enerji alışverişlerini kontrol etmektedir ve rüzgar çiftliğinden kaynaklı olası güç üretim tahminleri sapmalarını önlemek için gerekirse öncelikle kendi bünyesindeki pompalanmış su tabanlı enerji depolama ünitesinden faydalanmaktadır. Santralin üretimi herhangi bir zaman diliminde istenmeyen şekilde ilgili saat için planlanan güç üretim değerinden aşağıda veya yukarıda olursa, fazla olan enerji dengeleme piyasasına satılmaktadır veya eksik kalan enerji dengeleme piyasasından satın alınmaktadır.

Rüzgar çiftliğine ait güç üretiminin elde edilmesi amacıyla Oakland, Kaliforniya bölgesinden alınan ve Bölüm 4.1.2'de ayrıntılı olarak tanımlanan gerçek rüzgar hızları kullanılmıştır. Alınan hız değerleri gerçek bir türbine ait güç eğrisi yardımıyla güç değerlerine dönüştürülmüştür. Pompalanmış su tabanlı enerji depolama ünitesi modelinde ise temel olarak (Morales vd., 2013)'te verilen enerji depolama ünitesi modeli göz önüne alınmıştır ve bu model üzerinde önerilen çalışmadaki hedeflere ulaşmak amacıyla çeşitli değişiklikler yapılmıştır. Ayrıca, üst rezervdeki ve alt rezervdeki su hacmi miktarlarının değişimini zamana bağlı olarak ifade eden denklemlerde, buharlaşma ve yağmur gibi dış etmen tabanlı hacim artışı ve azalışı etkileri hesaba katılmıştır.

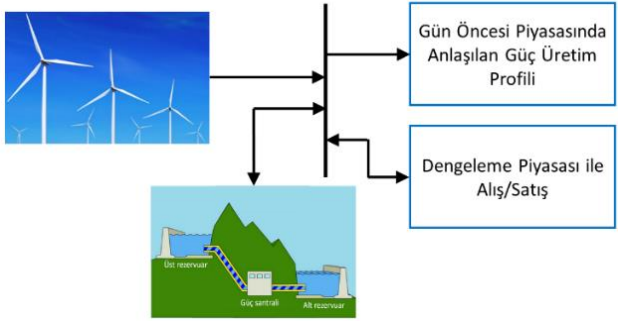

**Şekil 1. Önerilen yapıya ait temel şema.**

Önerilen sistemde amaç, güç santrali sahibinin saatlik ilgili fiyatlar ile anlaşılan gün öncesi piyasasındaki ve her 15 dakika için farklı alış/satış fiyatı olan dengeleme piyasasındaki işletimi sonunda elde edeceği günlük kazancı maksimize etmektir. Bu amaç aynı zamanda enerji maliyetlerinin yüksek olduğu tepe zamanlardaki toplam yük talebini de azaltacaktır.

### 3.2. Matematiksel Formülasyon

Güç santrali sahibi açısından maksimum kazanç ve sistem işletmecisi açısından azaltılmış tepe yük talebi hedeflerini gerçekleştirebilmek amacıyla önerilen optimizasyon problemine ait amaç fonksiyonu ve kısıtlar aşağıda alt başlıklar halinde verilmektedir.

### 3.2.1. Amaç Fonksiyonu

Belirtildiği üzere, güç santralinin sağlaması gereken ve 36 saat öncesinde güç sistemi operatörü ile birlikte belirlenen bir enerji talebi bulunmaktadır. Bu talebin sağlanamaması durumlarında şebekeden alınacak ve şebekeye verilecek olan enerjinin fiyatı dengeleme piyasasına ait fiyatlar üzerinden belirlenecektir. Denklem (1)'de her iki durumda güç sistemi sahibinin gün boyunca ve farklı fiyatlandırma senaryoları için elde edebileceği TL cinsinden toplam kazanç gösterilmektedir.

$$\text{maks. } Kazanç = \sum_s \sum_t \left( P_t^{planlanan,göp} \cdot \lambda_t^{göp} \cdot \Delta T_{göp} \right) + \left( P_{s,t}^{satılan,dp} \cdot \lambda_{s,t}^{satılan,dp} - P_{s,t}^{alınan,dp} \cdot \lambda_{s,t}^{alınan,dp} \right) \cdot \Delta T_{dp} \quad (1)$$

### 3.2.2. Güç Dengesi

Önerilen güç santrali yapısı içerisinde üretim ve tüketim arasındaki güç dengesi Denklem (2) ile sağlanmaktadır. Denklem (2)'den görülebileceği üzere üretimindeki belirsizlikleri göz önüne alabilmek amacıyla rüzgar gücü için de farklı senaryolar eşitliğe dahil edilmiştir.

$$P_{s,t}^{rüzgar} + P_{s,t}^{alınan,dp} + P_{s,t}^{salınan,ps} = P_t^{planlanan,göp} + P_{s,t}^{satılan,dp} + P_{s,t}^{pompalanan,ps}, \forall t \quad (2)$$

### 3.2.3. Pompalanmış Su Tabanlı Enerji Depolama Ünitesinin Modeli

Pompalanmış su tabanlı enerji depolama ünitesi için (Morales vd., 2013)'te verilen ilgili denklemler temel alınarak oluşturulan model, rezervuarlardaki suyun o andaki ihtiyaca göre aşağı veya yukarı taşınması ve bu sayede güç üretilmesi veya depolanması prensibine dayanmaktadır. Burada, Denklem (3) t anında pompalanmış su tabanlı ünite tarafından üretilen gücü ve (4) t anında üst rezervuara su pompalamak için harcanan gücü tanımlamaktadır. Denklemler (5) ve (6) alt ve üst rezervuarlardaki su miktarlarının değişimini temsil etmektedir. Rezervuarlardaki su miktarının alt ve üst limitleri (7) ve (8) ile korunmaktadır. Denklemler (9) ve (10) ile rezervuarlarda ilk anda bulunan su miktarları tanımlanmaktadır. Su akışının maksimum değeri ise (11) ve (12) ile her iki çalışma modu için sınırlandırılmaktadır. Denklemler (13) ve (14) ile santralin aynı anda iki modda birden çalışmasının ve (15) ve (16) ile dengeleme piyasasında aynı anda alım ve satım işlemlerinin yapılmasının önüne geçilmektedir. Son olarak, (17) ve (18) pompalanmış su tabanlı ünitenin şebekeden güç çekerek üst rezervuara su pompalamasını, diğer bir ifadeyle üretici olarak kayıtlı olduğu halde iki piyasa arası arbitraj yaparak tüketici gibi davranmasını önlemektedir.

$$P_{s,t}^{salınan,ps} = \sigma^{salınan,ps} \cdot q_{s,t}^{salınan,ps}, \forall t \quad (3)$$

$$P_{s,t}^{pompalanan,ps} = \sigma^{pompalanan,ps} \cdot q_{s,t}^{pompalanan,ps}, \forall t \quad (4)$$

$$v_{s,t}^{ÜST,ps} = v_{s,t-1}^{ÜST,ps} + \left( q_{s,t}^{pompalanan,ps} - q_{s,t}^{salınan,ps} \right) \cdot \Delta T_{dp} + v_t^{yağmur} - v_t^{buhar}, \forall t > 1 \quad (5)$$

$$v_{s,t}^{ALT,ps} = v_{s,t-1}^{ALT,ps} + \left( q_{s,t}^{salınan,ps} - q_{s,t}^{pompalanan,ps} \right) \cdot \Delta T_{dp} + v_t^{yağmur} - v_t^{buhar}, \forall t > 1 \quad (6)$$

$$V^{ÜST,ps,min} \leq v_{s,t}^{ÜST,ps} \leq V^{ÜST,ps,maks}, \forall t \quad (7)$$

$$V^{ALT,ps,min} \leq v_{s,t}^{ALT,ps} \leq V^{ALT,ps,maks}, \forall t \quad (8)$$

$$v_{s,t}^{ÜST,ps} = V^{ÜST,ps,başl}, eğer\ t = 1 \quad (9)$$

$$v_{s,t}^{ALT,ps} = V^{ALT,ps,başl}, eğer\ t = 1 \quad (10)$$

$$q_{s,t}^{pompalanan,ps} \leq Q^{maks,ps}, \forall t \quad (11)$$

$$q_{s,t}^{salınan,ps} \leq Q^{maks,ps}, \forall t \tag{12}$$

$$P_{s,t}^{pompalanan,ps} \leq N \cdot (1 - u_{s,t}^{ps}), \forall t \tag{13}$$

$$P_{s,t}^{salınan,ps} \leq N \cdot u_{s,t}^{ps}, \forall t \tag{14}$$

$$P_{s,t}^{satılan,dp} \leq N \cdot u_{s,t}^{dp1}, \forall t \tag{15}$$

$$P_{s,t}^{alınan,dp} \leq N \cdot (1 - u_{s,t}^{dp1}), \forall t \tag{16}$$

$$P_{s,t}^{alınan,dp} \leq N \cdot u_{s,t}^{dp2}, \forall t \tag{17}$$

$$P_{s,t}^{pompalanan,ps} \leq N \cdot (1 - u_{s,t}^{dp2}), \forall t \tag{18}$$

## 4. Testler

Pompalanmış su tabanlı enerji depolama ünitesinin güç santralinin kazancı üzerindeki etkilerini inceleyebilmek amacıyla benzetim çalışmalarında kullanılan veriler ve farklı durumlar için gerçekleştirilen analizler aşağıda verilmektedir.

### 4.1. Kullanılan Veri

Benzetim çalışmalarında kullanılan talep edilen güce, rüzgar modeline, pompalanmış su tabanlı enerji depolama ünitesinin modeline ve enerjinin fiyatlandırılmasına ait veriler aşağıdaki alt başlıklarda ayrı ayrı verilmiştir.

#### 4.1.1. Gün Öncesi Piyasasında Belirlenen Güce Ait Veriler

Güç santralinden beklenen bir günlük üretim değerlerinin 36 saat önceden belirlendiği varsayılmıştır. Rüzgar çiftliğinin üretim kapasitesinden tam olarak faydalanabilmek amacıyla sistem işletmecisine bildirilen saatlik güç üretimi değerleri, rüzgar çiftliğinden beklenen gerçek değerler olarak düşünülmüştür. Bu amaçla, on farklı senaryoya ait rüzgar gücünün her bir zaman aralığı için ortalaması alınmıştır. Elde edilen değerler Şekil 2'de gösterilmektedir.

Bu sayede, her bir senaryo için rüzgar çiftliğinin çıkış gücünün bir günlük süre boyunca bazı zamanlarda talep edilen gücün üzerinde, bazı zamanlarda ise altında kalması sağlanmıştır. Üretim ve tüketim arasında farkların oluştuğu zamanlarda pompalanmış su tabanlı enerji depolama ünitesinin ihtiyaç fazlası gücü depo etmesi veya gerekli eksik gücü tamamlaması beklenmektedir.

#### 4.1.2. Rüzgar Çiftliği Modeline Ait Veriler

Benzetim çalışmalarında kullanılan rüzgar hızı verisi, Amerika'da bulunan çeşitli hava istasyonları ve havaalanlarından beşer dakikalık aralıklarla ölçülen havacılık amaçlı rutin hava raporu (METAR) verisidir (Iowa State University of Science and Technology, 2017). Benzetim çalışmalarında 15'er dakikalık veriler kullanıldığı için yüksek doğrulukta ölçümler içeren bu verilerden faydalanılmıştır. Geniş bir veri seti içerisinden, benzetim çalışmalarında farklı koşulları gözlemleyebilmek amacıyla oldukça değişken bir rüzgar hızı profiline sahip olan Oakland, Kaliforniya bölgesine ait veriler seçilmiştir.

Önceki bölümde bahsedildiği üzere benzetim çalışmalarında kullanılan rüzgar türbinlerine ait çıkış gücü, 3 MW'lık bir türbine ait güç eğrisi yardımıyla gerçek rüzgar hızlarından elde edilmiştir (Vestas Wind Systems A/S R&D Department, 2004). Literatürde bu yöntem yaygın bir şekilde kullanıldığından türbine ait güç eğrisi ve bu eğriye ait doğrusal modele bu bölümde yer verilmemiştir. Farklı durumlardaki güç santralinin çalışmasını inceleyebilmek amacıyla gerçek rüzgar hızlarından, rulet tekerliği (roulette wheel) yöntemine dayanan bir yaklaşım ile dokuz farklı rüzgar hızı elde edilmiştir ve daha sonra bu değerler bahsedilen yöntemle güç değerlerine çevrilmiştir. Şekil 3'te farklı karakteristiklere sahip olan gerçek değer dahil on adet senaryoya ait rüzgar güçlerinin betimsel istatistikleri gösterilmiştir. Şekilde gösterilen kutulara dik olan çizgiler, ilgili zaman için on adet senaryo içerisindeki maksimum ve minimum değerlerini göstermektedir. Kutunun ortasında çizgi senaryolara ait medyan değerini, altındaki ve üzerindeki çizgiler ise sırasıyla birinci ve üçüncü çeyreklikleri temsil etmektedir.

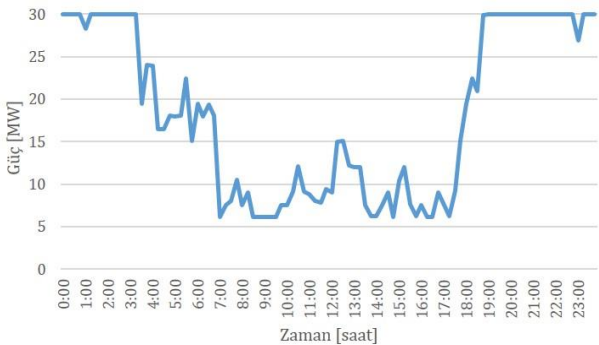

**Şekil 2. Gün öncesi piyasasında belirlenen ve güç santralinin üretmesi beklenen saatlik güç.**

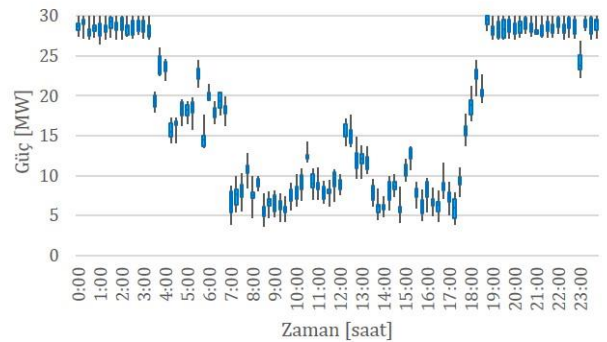

**Şekil 3. Farklı senaryolara ait rüzgar çiftliği çıkış gücünün istatistiksel gösterimi.**

### 4.1.3. Pompalanmış Su Tabanlı Enerji Depolama Ünitesinin Modeline Ait Veriler

Benzetim çalışmalarında kullanılan ve 3. bölümdeki denklemlerde yer alan pompalanmış su tabanlı enerji depolama ünitesine ait olan parametreler Tablo 1'de verilmiştir. Bu değerlerin seçimi için (Morales vd., 2013)'te verilen değerler baz alınmıştır. Rezervuarlara ait büyüklükler, göz önüne alınan rüzgar çiftliğinin gücü ile orantılı olması açısından % 25 daha büyük olarak alınırken, bu değerler dışındaki büyüklükler değiştirilmeden kullanılmıştır. Bir ön çalışma ile belirlenebilecek en uygun güç ve su pompalama kapasitesine sahip bir depolama ünitesinin kullanılması durumunda elde edilecek ekonomik ve teknik faydaların artacağı belirtilebilir.

**Tablo 1. Pompalanmış su tabanlı enerji depolama ünitesine ait parametreler (Morales vd., 2013).**

| Parametre | Değer | Birim | Parametre | Değer | Birim |
|---|---|---|---|---|---|
| $V^{ALT,ps,başl}$ | 50 | Hm³ | $V^{ÜST,ps,başl}$ | 50 | Hm³ |
| $V^{ALT,ps,maks}$ | 100 | Hm³ | $V^{ÜST,ps,maks}$ | 100 | Hm³ |
| $V^{ALT,ps,min}$ | 10 | Hm³ | $V^{ÜST,ps,min}$ | 10 | Hm³ |
| $\sigma^{pompalanan,ps}$ | 1,2 | MW/(Hm³/h) | $\sigma^{salınan,ps}$ | 0,8 | MW/(Hm³/h) |
| $Q^{maks,ps}$ | 20 | Hm³/h | | | |

### 4.1.4. Enerji Fiyatlandırmasına Ait Veriler

Önceki bölümde belirtildiği gibi güç sistemindeki enerji alışverişlerinde gün öncesi piyasasındaki ve dengeleme piyasasındaki enerji fiyatları beraber dikkate alınmaktadır. Bu çalışmada gün öncesi piyasasındaki enerji fiyatlarının saatlik olarak önceden belirlendiği varsayılmıştır. Benzetim çalışmalarında kullanılan ve Şekil 4'te gösterilen enerji fiyatları, Enerji Piyasaları İşletme Anonim Şirketi (EPİAŞ) internet sitesinden alınan gerçek değerlerdir.

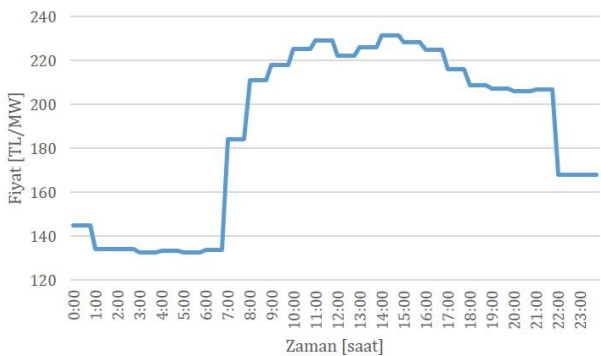

**Şekil 4. Gün öncesi piyasasında belirlenen saatlik enerji fiyatı.**

Dengeleme piyasasındaki fiyatlar pratikteki uygulamalarda güç sistemindeki üretim-tüketim dengesine göre sürekli olarak değişmektedir. Güç sistemindeki diğer yükler ile ilgili bir bilgi bulunmadığından çalışma kapsamında rüzgar gücüne ait her bir senaryo için farklı bir dengeleme piyasası fiyat değişimi göz önüne alınmıştır. Şekil 5'te benzetim çalışmalarında kullanılan ve her bir 15 dakikalık periyot için tüm senaryolara ait olan betimsel istatistikler verilmiştir.

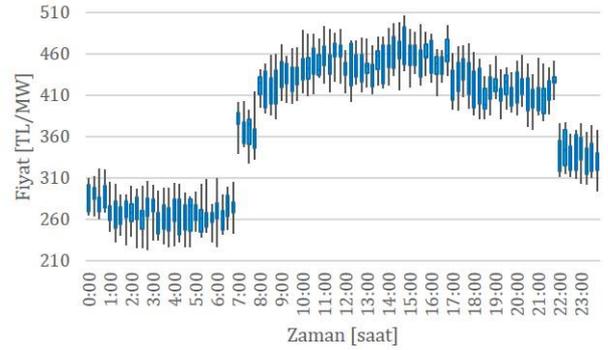

**Şekil 5. Farklı senaryolara ait 15 dakikalık dengeleme piyasası enerji alış-satış fiyatlarının istatistiksel gösterimi.**

### 4.2. Benzetim Çalışmaları ve Sonuçlar

Önerilen optimizasyon probleminin etkinliğini değerlendirebilmek amacıyla Bölüm 3.2'de verilen model, Generic Algebraic Modeling System (GAMS) yazılımı ortamında bir karma tam sayılı programlama (mixed-integer programming – MIP) problemi olarak kodlanmıştır. Optimizasyon problemi odaklı bir yazılım olan GAMS literatürde özellikle son yıllarda yaygın bir şekilde kullanılmaktadır (Amrollahi ve Bathaee, 2017; Nojavan vd., 2018). Problemin çözümü içinse IBM tarafından Simpleks optimizasyon çözüm yöntemi temel alınarak geliştirilmiş olan ve özellikle doğrusal optimizasyon problemlerinin çözümünde oldukça yüksek bir performans gösteren CPLEX (IBM, 2018) çözücüsü kullanılmıştır.

Pompalanmış su tabanlı enerji depolama ünitesini farklı koşullar altında test edebilmek ve elde edilen sonuçları değerlendirebilmek amacıyla benzetim çalışmalarında Tablo 2'den görülebileceği gibi dört farklı durum ele alınmıştır. İlk durumda güç santralinin yalnızca rüzgar çiftliğinden oluştuğu varsayılmıştır. İkinci durumda ise yine yalnızca rüzgar çiftliğinden oluşan güç santralinin, Şekil 2'de verilen gün öncesi piyasasında belirlenmiş yük profili yerine % 20 daha az güç değerlerine sahip bir yük profilini sağlayacağı düşünülmüştür. Böylece rüzgar santralinin bazı saatlerde sağlayamadığı yüksek güç değerleri nedeniyle dengeleme piyasasından daha az güç çekmesi ve kazancını arttırması beklenmektedir.

Pompalanmış su tabanlı ünitenin eklenmesi durumları ise Durumlar 3 ve 4'te incelenmiştir. İki durum arasındaki tek fark, üçüncü durumda depolama ünitesinin kullanımında dengeleme piyasasındaki değişken fiyatın öneminin dikkate alınmamasıdır. Bu sayede dördüncü durumda, önerilen optimizasyon yönteminin faydalarının gösterilmesi hedeflenmiştir.

**Tablo 2. Benzetim çalışmalarında göz önüne alınan farklı durumlara ait bilgiler.**

| Durum | Kaynak | Yük | Yöntem |
|---|---|---|---|
| 1 | Rüzgar çiftliği | Şekil 2'de verilen yük | Fiyat değişimini dikkate alan |
| 2 | Rüzgar çiftliği | Şekil 2'de verilen yükün % 80'i | Fiyat değişimini dikkate alan |
| 3 | Rüzgar çiftliği ve depolama ünitesi | Şekil 2'de verilen yük | Fiyat değişimini dikkate almayan |
| 4 | Rüzgar çiftliği ve depolama ünitesi | Şekil 2'de verilen yük | Fiyat değişimini dikkate alan |

Güç santralinden beklenen enerji üretimi ve rüzgar çiftliğinin çıkış gücü parametreler olarak kullanıldığından, dört durum için ortak olarak karşılaştırılabilecek değişkenler; dengeleme piyasasından satın alınan ve piyasaya satılan güç değerleridir. Şekil 6-9 bir günlük süre boyunca yalnızca rüzgar çiftliğinden veya rüzgar çiftliği-depolama ünitesinden oluşan güç santralinin dengeleme piyasası ile olan enerji alışverişlerinin on farklı senaryo için ortalamasını göstermektedir.

Şekil 6'dan görülebileceği üzere birinci durumda rüzgar çiftliği istenen enerji değerini karşılayamadığı zamanlarda dengeleme piyasasından eksik güç miktarını temin etmekte veya ihtiyaç fazlası enerji üretildiğinde bu enerjiyi dengeleme piyasasına satmaktadır. Gün öncesi piyasasına sağlanacak enerji değeri ile rüzgar çiftliğinin ürettiği enerji değeri her saat için birbirine yakın değerler olarak seçildiğinden ve güç üretimi farklı senaryolarda çoğunlukla şebekeye verilen güç değerinin altında kaldığından, birinci durumda genellikle dengeleme piyasasından enerji satın alınmaktadır.

İkinci durumda kazancı arttırabilmek amacıyla Şekil 2'de verilen saatlik yük değişimi % 20 azaltılarak, farklı senaryolara ait üretim değerlerindeki belirsizlikler nedeniyle dengeleme piyasasından alınan enerjinin sınırlanması amaçlanmıştır. Böylece, Şekil 7'den görülebileceği üzere ikinci durumda dengeleme piyasasına yapılan enerji satışı ilk duruma göre önemli derecede artmıştır ve enerji alımı önemli derecede azalmıştır. Sonuç olarak ikinci durumda ekonomik kazancın artacağı Şekil 7'den açıkça görülmektedir. Kazançtaki bu artışın en önemli sebebi, dengeleme piyasası ile enerji alışverişinde herhangi bir kısıtlama bulunmamasıdır. Dengeleme piyasasında enerji alımlarının yalnızca ihtiyaç duyulan belli sürelerde gerçekleştirilmesi durumunda, rüzgar çiftliği üretimindeki beklenen değerlerin üzerindeki enerji miktarları rüzgar türbini kontrol yöntemleriyle sınırlandırılmalıdır. Aksi halde uygulanacak cezalar ile kazanç değerleri önemli ölçüde azalacaktır.

Enerji fiyatı değişiminin optimizasyon problemi içerisinde dikkate alındığı ve alınmadığı durumlara ait dengeleme piyasası ile olan enerji alışverişleri Şekiller 8 ve 9'da verilmiştir. Şekil 8'den görülebileceği üzere üçüncü durumda güç sisteminden çekilen enerji miktarı oldukça sınırlı kalmıştır. Bunun nedeni, üçüncü durumda uygulanan amaç fonksiyonunun dengeleme piyasası enerji fiyatlarını dikkate almaması sebebiyle ileride daha yüksek fiyatlı periyotlarda enerji satabilmek amacıyla düşük fiyatlı periyotlarda enerji satın alınmamasıdır. Dördüncü durumda ise dengeleme piyasasında fiyatın düşük olduğu zamanlarda pompalanmış su tabanlı ünitede enerji depolanmaktadır ve depolanan enerji fiyatın yüksek olduğu zamanlarda şebekeye satılmaktadır. Bu sayede en yüksek kazanca ulaşılması sağlanmaktadır.

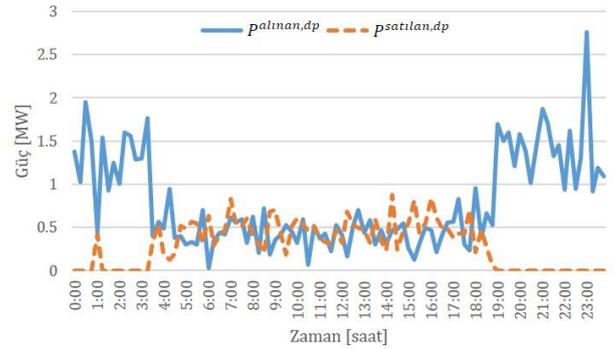

**Şekil 6. Birinci durum için dengeleme piyasası ile yapılan enerji alışverişleri.**

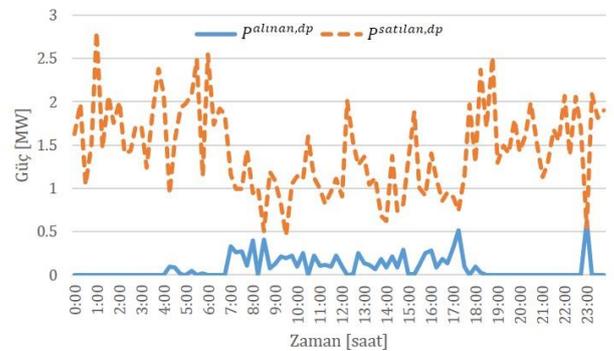

**Şekil 7. İkinci durum için dengeleme piyasası ile yapılan enerji alışverişleri.**

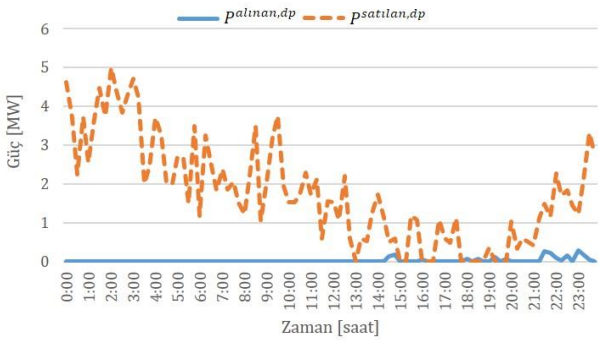

**Şekil 8.** Üçüncü durum için dengeleme piyasası ile yapılan enerji alışverişleri.

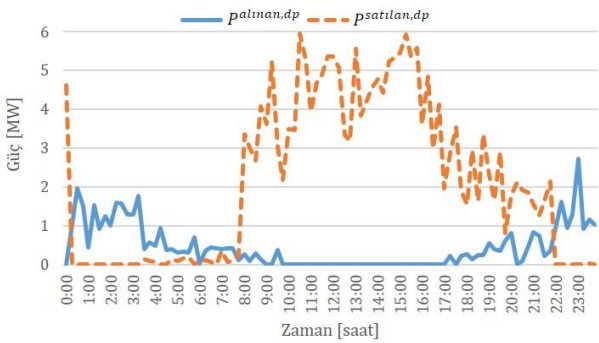

**Şekil 9.** Dördüncü durum için dengeleme piyasası ile yapılan enerji alışverişleri.

Pompalanmış su tabanlı enerji depolama ünitesinin su pompalama ve türbin modlarındaki güç değerleri Durumlar 3 ve 4 için sırasıyla Şekiller 10 ve 11'de gösterilmektedir. Şekil 10'dan görülebileceği üzere depolanmış enerjinin büyük bir bölümü enerji fiyatının daha düşük olduğu zamanlarda kullanılmıştır. Dördüncü durumda ise aksine enerji fiyatının düşük olduğu periyotlarda korunan enerji yalnızca yüksek fiyatlı zamanlarda harcanmıştır.

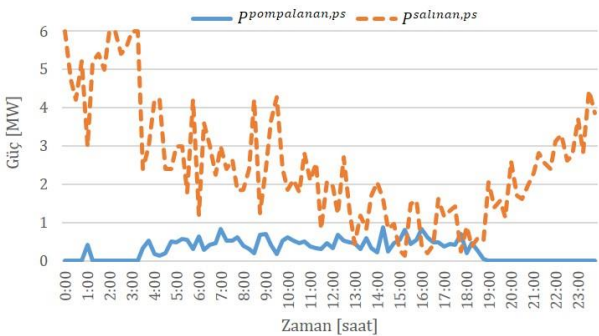

**Şekil 10.** Üçüncü durum için pompa ve türbin modundaki güçler.

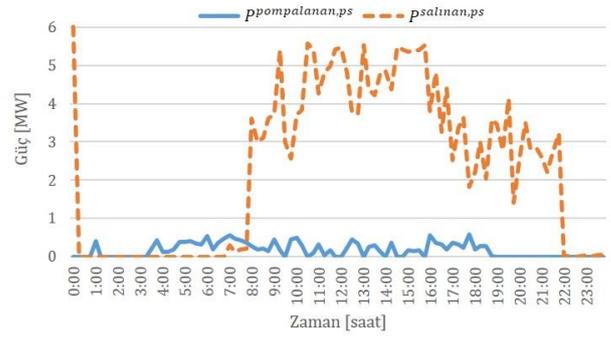

**Şekil 11.** Dördüncü durum için pompa ve türbin modundaki güçler.

Şekiller 12 ve 13'te gösterilen Durumlar 3 ve 4'e ait rezervuarlardaki su miktarları değişimi de Şekiller 10 ve 11 için belirtilen hususları doğrulamaktadır. Dengeleme piyasasındaki enerji fiyatı değişiminin dikkate alınmadığı Durum 3'te depolanmış enerji her bir periyotta kullanılırken Durum 4'te öncelikle fiyat düşükken belli bir miktar daha enerji depolanmış ve fiyat yükseldikten sonra enerjinin tamamı kullanılmıştır.

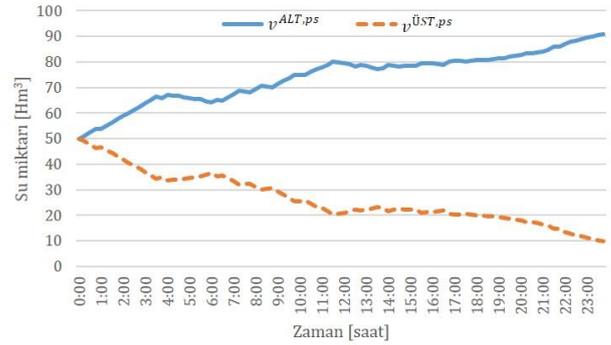

**Şekil 12.** Üçüncü durum için alt ve üst rezervuarlardaki su miktarı değişimleri.

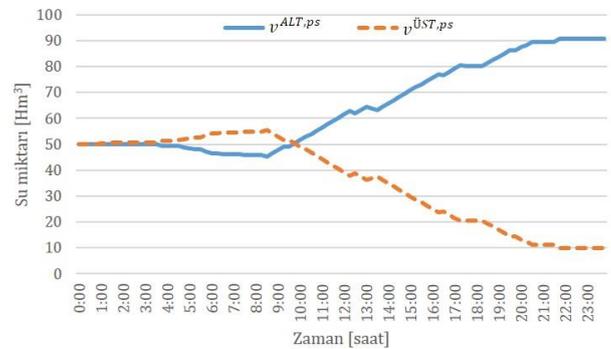

**Şekil 1.** Dördüncü durum için alt ve üst rezervuarlardaki su miktarı değişimleri.

Tüm durumlar için benzetim çalışmaları sonucunda elde edilen dengeleme piyasasındaki enerji alışverişleri, enerji alışverişleri sonucunda ortaya çıkan kazanç değerleri ve ilk duruma göre kazançtaki artış değerleri Tablo 3'te verilmiştir. Tablo 3'te verilen kazanç değerleri on adet senaryo için bir gün içerisindeki kazanç değerlerinin ortalamasını göstermektedir. Gün öncesi piyasasına ait güç değerlerinin Tablo 3'te yer almamasının nedeni, bu değerlerin tüm durumlar için (% 20 oranında azaltılmış Durum 2 hariç) aynı olmasıdır. Tablodaki değerler, pompalanmış su tabanlı depolama ünitesi kullanımının ve önerilen fiyat değişimi tabanlı optimizasyon probleminin diğer durumlara göre çok daha olumlu sonuçlar verdiğini göstermektedir.

**Tablo 1. Farklı durumlarda satın alınan/satılan güç ve elde edilen kazanç değerleri.**

| Durum | Ortalama Satın Alınan Güç [MW] | Ortalama Satılan Güç [MW] | Kazanç [TL] | Kazanç Artışı [%] |
|---|---|---|---|---|
| 1 | 743,61 | 287,58 | 73649,39 | - |
| 2 | 82,86 | 1381,21 | 81386,40 | 10,51 |
| 3 | 19,42 | 1705,82 | 91619,25 | 24,40 |
| 4 | 429,74 | 2053,67 | 96928,97 | 31,61 |

Tablo 3'de verilen kazanç değerleri ve benzetim çalışmalarında elde edilen sonuçları gösteren tüm şekiller dikkate alındığında dikkat edilmesi gereken iki önemli husus bulunmaktadır. Bunlardan ilki, farklı durumlar için verilen dengeleme piyasasındaki enerji alış ve satışlarında herhangi bir kısıtlama bulunmamaktadır. Başka bir ifadeyle, her ihtiyaç olduğunda şebekeden istenen miktarda enerji satın alınabilmektedir veya şebekeye istenen miktarda enerji satılabilmektedir. Ancak bu durum pratikte genellikle mümkün olmamaktadır. Gerçek uygulamalarda, yalnızca şebekedeki güç santrallerinin 36 saat önceden belirttikleri enerji değerlerini sağlayamadıkları durumlarda diğer santrallerden dengeleme piyasası fiyatları üzerinden bir enerji alımı gerçekleştirilmektedir. Belirtilen husus, benzetim çalışmalarında Durumlar 1 ve 2'ye kazanç değerleri açısından bir avantaj sağlamaktadır. Bunun nedeni, gerçek bir uygulamada gün öncesi piyasasında belirlenen saatlik güç değerlerinden sapmalar olması durumunda bir rüzgar santralinin ceza ödemesi gerekirken, Durumlar 1 ve 2'de belirlenen güç fazlası, dengeleme piyasası fiyatı üzerinden şebekeye satılmaktadır. Bu nedenle, elde edilen değerler özellikle Durum 2 için olması gerekenden daha fazla çıkmıştır. Benzer şekilde, Durumlar 3 ve 4 için üst rezervuardaki su miktarının ilk değerinin 50 Hm³ ve minimum değerinin 10 Hm³ olarak alınması, ilk iki durumla yapılan karşılaştırmalarda maddi kazançlar açısından bir avantaj sağlamaktadır.

Belirtilen iki problemi çözerek daha gerçekçi bir analiz yapabilmek için öncelikle bölgedeki rüzgar potansiyeline ve gün içerisindeki üretim-tüketim dengesine göre her iki ünite için en uygun kurulu güç değerlerinin belirlenmesi gerekmektedir. Ayrıca, belirleme hesaplarında, ilk yatırım maliyeti, arazinin durumu, vb. gibi çeşitli ekonomik ve teknik kısıtların da oldukça önemli bir etkisi olacaktır. Daha sonra, rüzgar çiftliğinin her bir zaman adımında üretebileceği en yüksek güç değerlerinin belirlenmesi ve rüzgar gücündeki belirsizlikler de hesaba katılarak belirlenen en yüksek güç değerlerine göre gün öncesi piyasasındaki üretim profilinin oluşturulması gerekmektedir. Ancak bu çalışmanın temel amacı, pompalanmış su tabanlı enerji depolama ünitelerinin değişken enerji fiyatına sahip güç sistemlerindeki olumlu etkilerini incelemek olduğu için belirtilen işlemler ve analizler çalışmanın kapsamı dışında kalmaktadır. Ayrıca, dengeleme piyasasındaki ihtiyaçlar ve enerji fiyatları tamamen o andaki koşullara göre bir sonraki periyod için belirlendiğinden, tam olarak gerçekçi bir benzetim çalışması gerçekleştirmek mümkün değildir. Bu nedenle, yukarıda belirtilen varsayımlar kabul edilerek bu çalışmada yalnızca temel amaca yönelik analizlere odaklanılmıştır.

Belirtilen şartlara rağmen yine de farklı koşullardaki sonuçlar hakkında bir bilgi edinebilmek amacıyla optimizasyon yöntemine üç yeni kısıt eklenerek benzetim çalışmaları tekrarlanmıştır. Burada ilk iki kısıt ile her bir zaman aralığında güç sistemine satılabilecek güç miktarı, 9:00-21:00 saatleri arasında 2 MW, diğer zamanlarda ise 0,5 MW ile sınırlandırılmıştır. Belirtilen zaman aralıkları, Şekil 4 ve Şekil 5'teki enerji fiyatı bilgileri göz önüne alınarak pratikte gerçekleşmesi muhtemel bir durum elde edebilmek amacıyla seçilmiştir. Üçüncü kısıt ise alt rezervuardaki su miktarının bir günlük benzetim çalışması sonunda ilk andaki değeri ile aynı olmasını sağlamak için kullanılmıştır. Yeni kısıtların uygulanması ile elde edilen sonuçlar Durum 3 ve Durum 4 için Şekil 14-19'da gösterilmiştir. Yalnızca Durum 3 ve Durum 4'e ait sonuçların verilmesinin nedeni, önceki koşullarda bu iki durumun en etkin sonuçları sağlamış olmasıdır.

Oluşturulan yeni koşullar altında elde edilen kazanç değerleri ise, Durum 3 ve Durum 4 için sırasıyla 74779,34 TL ve 76022,17 TL olarak hesaplanmıştır. Yeni kısıtların optimizasyon problemi içerisine dahil edilmesi sonucunda elde edilen kazanç değerlerinden ve Şekil 14-19'dan görülebileceği üzere, eklenen kısıtlar neticesinde, farklı güç değerlerinin ve rezervuarlardaki su miktarlarının zamanla değişimlerinde önemli farklılıklar oluşmuştur. Algoritmaya dahil edilen kısıtlar, enerji alışverişlerini belli bir ölçüde sınırlandırdıkları için şebeke ile güç değişimleri Şekiller 14 ve 15'ten görülebileceği şekilde önemli ölçüde azalmıştır. Belirtilen değişimler, Şekiller 16 ve 17'de açıkça görüldüğü gibi pompalanan ve serbest bırakılan su miktarlarını da doğrudan etkilemişlerdir.

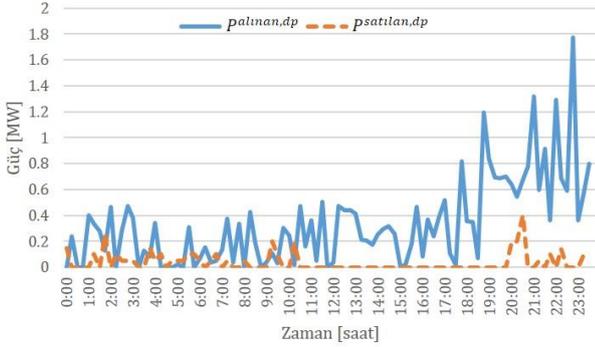

**Şekil 2.** Yeni kısıtlar altında üçüncü durum için dengeleme piyasası ile yapılan enerji alışverişleri.

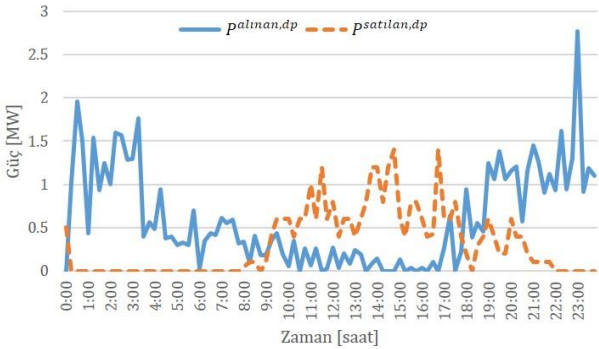

**Şekil 3.** Yeni kısıtlar altında dördüncü durum için dengeleme piyasası ile yapılan enerji alışverişleri.

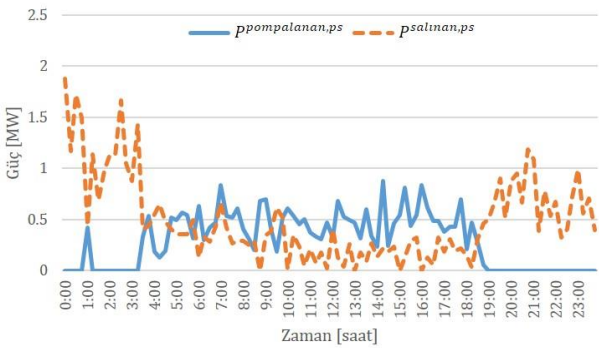

**Şekil 4.** Yeni kısıtlar altında üçüncü durum için pompa ve türbin modundaki güçler.

Son olarak, Şekiller 18 ve 19 göz önüne alındığında, üçüncü kısıtın gerektirdiği şartı sağlayabilmek adına pompa depolamalı ünitenin ilk örneğe kıyasla daha sınırlı ölçekte kullanıldığı sonucuna ulaşılabilir. Şekil 18 ve 19'da rezervuarlardaki su miktarlarının son durumlarında baştaki değerlerinden sapmalarının nedeni, model yapısına dahil edilen yağmur ve buharlaşma etkenleridir. Yeni kısıtların dahil edilmesi sonucunda ortaya çıkan tüm bu farklılıklara rağmen, önerilen optimizasyon probleminin yeni kısıtlar eklenmeden önceki sonuçlar ile benzer şekilde, Durum 3'e göre daha yüksek kazanç sağladığı yukarıda verilen değerlerden anlaşılmaktadır.

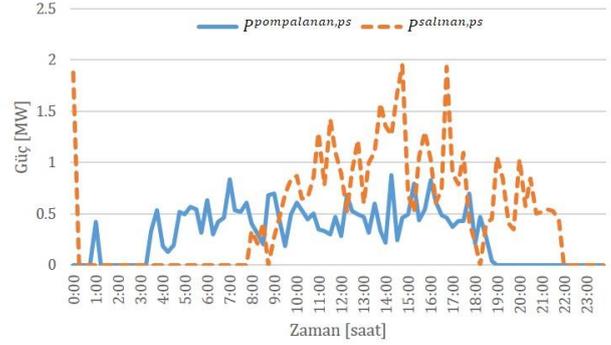

**Şekil 5.** Yeni kısıtlar altında dördüncü durum için pompa ve türbin modundaki güçler.

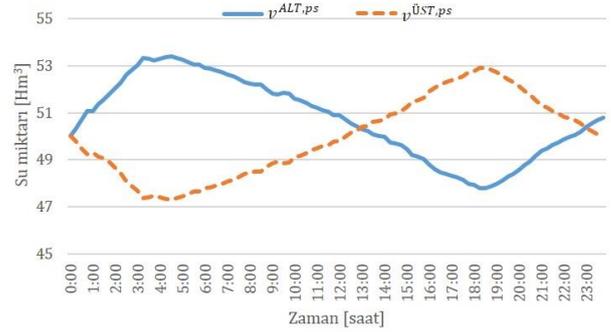

**Şekil 6.** Yeni kısıtlar altında üçüncü durum için alt ve üst rezervuarlardaki su miktarı değişimleri.

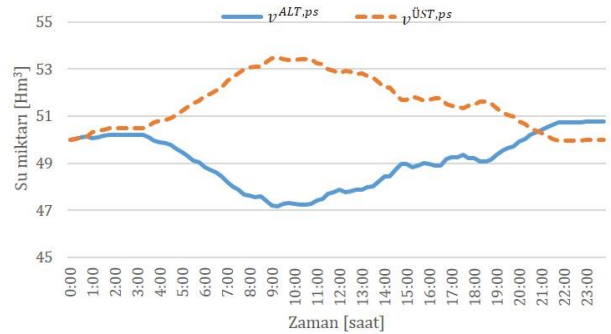

**Şekil 7.** Yeni kısıtlar altında dördüncü durum için alt ve üst rezervuarlardaki su miktarı değişimleri.

Enerji depolama ünitelerinin enerji maliyeti üzerindeki olumlu etkilerine ek olarak güç sistemindeki tepe yüklerin azaltılmasında da önemli bir rolü bulunmaktadır. Temel olarak, enerjinin fiyatının yüksek olduğu zamanlar enerji talebinin de yüksek olduğu zamanlara denk geldiği için kazanç değerlerini arttırmak tepe yük talebini de azaltacaktır. Şekiller 4 ve 5'e göre şebekedeki yük 13:00 ile 17:00 saatleri arasında tepe değerlerine ulaşmaktadır. Bu saatler arasında depolama ünitesinin şebekeye sağladığı 15 dakikalık güç miktarları, depolama sistemi içeren her iki durum için Şekil 20'de gösterilmiştir. Belirtilen zaman aralığı boyunca şebekeye verilen enerji miktarları ise Durumlar 3 ve 4 için sırasıyla 24,86 MWh ve 194,59 MWh olarak hesaplanmıştır. Özellikle Durum 4 için şebekeye sağlanan oldukça yüksek enerji değeri, şebekenin üretim ve tüketim arasındaki dengeyi sağlamasında yaşanması muhtemel ekonomik ve teknik sorunları önemli ölçüde azaltacaktır.

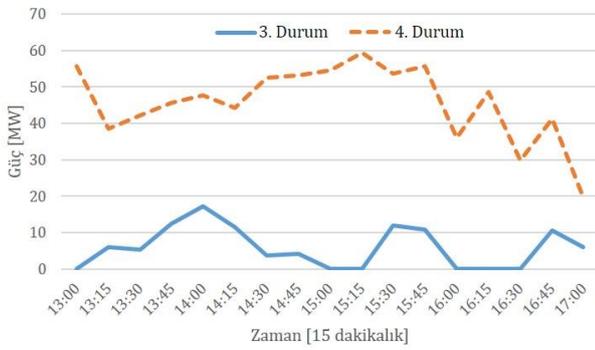

**Şekil 20.** Üçüncü ve dördüncü durumlar için tepe talep zamanlarında güç sistemine sağlanan güçler.

## 5. Sonuçlar

Bu çalışmada, gün öncesi ve dengeleme piyasalarının bulunduğu bir güç sisteminde, pompalanmış su tabanlı bir enerji depolama ünitesinin bir rüzgar çiftliği ile birlikte kullanılması durumunda güç santrali sahibine sağlayacağı ekonomik kazanç incelenmiştir. Bu kazancı arttırmak amacıyla önerilen bir optimizasyon probleminin etkinliği, farklı durumlar için elde edilen sonuçlarla yapılan karşılaştırmalar ile test edilmiştir. Gerçek rüzgar hızı ölçümlerine ve enerji fiyatı verilerine ek olarak bu değerlere belirsizlikler eklemek amacıyla oluşturulan farklı senaryoların kullanılmasıyla yapılan benzetim çalışmaları sonucunda pompalanmış su tabanlı ünitenin; yüksek güç değerlerini depolamakta yeterli olduğu, gerektiği zamanlarda şebekedeki eksik güç değerlerini sağlayabildiği ve enerjinin saatlik ve 15 dakikalık piyasa fiyatlarının göz önüne alınması durumunda önemli bir kazanç elde etmeye olanak sağladığı görülmüştür. Ayrıca, önerilen depolama sistemi ve değişken fiyat tabanlı optimizasyon probleminin şebekelerdeki en önemli sorunlardan biri olan tepe yük talebini de azaltabildiği gösterilmiştir.

Önerilen yaklaşım, sıkıştırılmış hava tabanlı enerji depolama üniteleri gibi farklı depolama sistemleri ile birlikte ve/veya farklı fiyatlandırma stratejilerinin kullanıldığı güç sistemlerinde benzer olumlu sonuçlar elde edilebilecek şekilde kullanılmaya uygundur. Güç sistemlerinde bu ünitelerden birden fazla sayıda bulunması durumundaki optimal enerji alışverişleri yazarlar tarafından yakın gelecekteki bir çalışma olarak planlanmaktadır. Benzer şekilde, önerilen yapının talep cevabı programları ile birlikte kullanılması durumunda, özellikle güç sistemi tarafında elde edilebilecek faydaların araştırılması da literatüre önemli katkılar sağlayabilir.

### Çıkar Çatışması

Yazarlar tarafından herhangi bir çıkar çatışması beyan edilmemiştir.

### Kaynaklar

Al-Swaiti, M.S., Al-Awami, A.T. ve Khalid, M.W. (2017). Co-optimized trading of wind-thermal-pumped storage system in energy and regulation markets, *Energy*, 138, 991-1005.

Amrollahi, M.H. ve Bathaee, S.M.T. (2017). Techno-economic optimization of hybrid photovoltaic/wind generation together with energy storage system in a stand-alone micro-grid subjected to demand response, *Applied Energy*, 202, 66-77.

Attya, A.B.T. ve Hartkopf, T. (2015). Utilising stored wind energy by hydro-pumped storage to provide frequency support at high levels of wind energy penetration, *IET Generation, Transmission & Distribution*, 9(12), 1485-1497.

Bruninx, K., Dvorkin, Y., Delarue, E., Pandžić, H., D'haeseleer, W. ve Kirschen. D.S. (2016). Coupling pumped hydro energy storage with unit commitment, *IEEE Transactions on Sustainable Energy*, 7(2), 786-796.

Chen, C.L., Chen, H.C. ve Lee, J.Y. (2016). Application of a generic superstructure-based formulation to the design of wind-pumped-storage hybrid systems on remote islands, *Energy Conversion and Management*, 111, 339-351.

de Boer, H.S., Grond, L., Moll, H. ve Benders, R. (2014). The application of power-to-gas, pumped hydro storage and compressed air energy storage in an electricity system at different wind power penetration levels, *Energy*, 72, 360-370.

de la Nieta, A.A.S., Contreras, J. ve Catalão, J.P. (2016). Optimal single wind hydro-pump storage bidding in day-ahead markets including bilateral contracts, *IEEE Transactions on Sustainable Energy*, 7(3), 1284-1294.

Elma, O., Taşcıkaraoğlu, A., İnce, A.T. ve Selamoğulları, U.S. (2017). Implementation of a dynamic energy management system using real time pricing and local renewable energy generation forecasts, *Energy*, 134, 206-220.

Foley, A.M., Leahy, P.G., Li, K., McKeogh, E.J. ve Morrison, A.P. (2015). A long-term analysis of pumped hydro storage to firm wind power, *Applied Energy*, 137, 638-648.

Global Wind Energy Council. (2016). Global wind report annual market update.

IBM. CPLEX çözücüsü. https://www.ibm.com/analytics/data-science/ prescriptive-analytics/cplex-optimizer]. Erişim tarihi: 19 Şubat 2018.

Iowa State University of Science and Technology. The Iowa Environmental Mesonet (IEM). http://mesonet.agron.iastate.edu/request/download.phtml?network=CA_ASOS. Erişim tarihi: 1 Ağustos 2017.

Khodayar, M.E., Shahidehpour, M. ve Wu, L. (2013). Enhancing the dispatchability of variable wind generation by coordination with pumped-storage hydro units in stochastic power systems, *IEEE Transactions on Power Systems*, 28(3), 2808-2818.

Kiran, B.D.H. ve Kumari, M.S. (2016). Demand response and pumped hydro storage scheduling for balancing wind power uncertainties: A probabilistic unit commitment approach, *International Journal of Electrical Power & Energy Systems*, 81, 114-122.

Kusakana, K. (2016). Optimal scheduling for distributed hybrid system with pumped hydro storage. *Energy Conversion and Management*, 111, 253-260.

Ma, T., Yang, H. ve Lu, L. (2014). Feasibility study and economic analysis of pumped hydro storage and battery storage for a renewable energy powered island, *Energy Conversion and Management*, 79, 387-397.

Ma, T., Yang, H., Lu, L. ve Peng, J. (2015a). Optimal design of an autonomous solar–wind-pumped storage power supply system, *Applied Energy*, 160, 728-736.

Ma, T., Yang, H., Lu, L. ve Peng, J. (2015b). Pumped storage-based standalone photovoltaic power generation system: Modeling and techno-economic optimization, *Applied Energy*, 137, 649-659.

Ming, Z., Kun, Z. ve Liang, W. (2014). Study on unit commitment problem considering wind power and pumped hydro energy storage, *International Journal of Electrical Power & Energy Systems*, 63, 91-96.

Morales, J.M., Conejo, A.J., Madsen, H., Pinson, P. ve Zugno, M. Integrating renewables in electricity markets: operational problems, Yayın No: 205, *Springer Science & Business Media*, USA, Springer, 2013.

Nojavan, S., Najafi-Ghalelou, A., Majidi, M. ve Zare, K. (2018). Optimal bidding and offering strategies of merchant compressed air energy storage in deregulated electricity market using robust optimization approach, *Energy*, 142, 250-257.

Papaefthymiou, S.V., Lakiotis, V.G., Margaris, I.D. ve Papathanassiou, S.A. (2015). Dynamic analysis of island systems with wind-pumped-storage hybrid power stations, *Renewable Energy*, 74, 544-554.

Papaefthymiou, S.V. ve Papathanassiou, S.A. (2014). Optimum sizing of wind-pumped-storage hybrid power stations in island systems, *Renewable Energy*, 64, 187-196.

Paterakis, N.G., Taşcıkaraoğlu, A., Erdinc, O., Bakirtzis, A.G. ve Catalao, J.P. (2016). Assessment of demand-response-driven load pattern elasticity using a combined approach for smart households, *IEEE Transactions on Industrial Informatics*, 12(4), 1529-1539.

Pérez-Díaz, J.I. ve Jiménez, J. (2016). Contribution of a pumped-storage hydropower plant to reduce the scheduling costs of an isolated power system with high wind power penetration, *Energy*, 109, 92-104.

Petrakopoulou, F., Robinson, A. ve Loizidou, M. (2016). Simulation and analysis of a stand-alone solar-wind and pumped-storage hydropower plant, *Energy*, 96, 676-683.

Sousa, J.A., Teixeira, F. ve Faias, S. (2014). Impact of a price-maker pumped storage hydro unit on the integration of wind energy in power systems, *Energy*, 69, 3-11.

Stoppato, A., Cavazzini, G., Ardizzon, G. ve Rossetti, A. (2014). A PSO (particle swarm optimization)-based model for the optimal management of a small PV (Photovoltaic)-pump hydro energy storage in a rural dry area, *Energy*, 76, 168-174.

Tascikaraoglu A. (2018a). Economic and operational benefits of energy storage sharing for a neighborhood of prosumers in a dynamic pricing environment, *Sustainable Cities and Society*, 38, 219-229.

Tascikaraoglu A. (2018b). Evaluation of spatio-temporal forecasting methods in various smart city applications, *Renewable and Sustainable Energy Reviews*, 82, 424-435.

Vestas Wind Systems A/S R&D Department. (2004). General Specification V90 – 3.0 MW. http://www.gov.pe.ca/photos/sites/envengfor/file/950010R1_V90-GeneralSpecification.pdf. Erişim tarihi: 19 Şubat 2018.

Vieira, B., Viana, A., Matos, M. ve Pedroso, J.P. (2016). A multiple criteria utility-based approach for unit commitment with wind power and pumped storage hydro, *Electric Power Systems Research*, 131, 244-254.

Yıldız, C. ve Şekkeli, M. (2016). Türkiye gün öncesi elektrik piyasasında rüzgar enerjisi ve pompaj depolamalı hidroelektrik santral için optimum teklif oluşturulması, *Pamukkale Üniversitesi Mühendislik Bilimleri Dergisi*, 22(5), 361-366.